\documentclass[aps, superscriptaddress, floatfix]{revtex4}

\usepackage{graphicx}

\begin{document}



\newcommand{\Dirac}{\rlap{\hspace{-.5mm} \slash} D}

\title{
 Light Quark Condensates at Nonzero Chemical Potentials}
\author{D.~Toublan}
\affiliation {Physics Department, University of Maryland, College
Park, MD 20742}

\date{\today}

\begin{abstract}
We show that the quark condensates for the two light up and down
flavors can have significantly different values in the hadronic
phase at nonzero temperature, baryon and isospin chemical
potentials.  We quantify this difference using a simple model.
\end{abstract}

\maketitle

\section{Introduction}
In recent years, an intense experimental and theoretical research
effort has been devoted to the study of the strong interaction at
nonzero temperature and density.  These studies are important to get
a better understanding of very different systems such as neutron
stars, heavy-ion-collision experiments, or the early universe.
 Usually, theoretical studies are performed at nonzero baryon
chemical potential, $\mu_B$, and zero isospin chemical potential,
$\mu_I$ \cite{cscRev, lattMuReview}. However, physical systems often
have both nonzero baryon and isospin chemical potentials. Prime
examples of such systems are neutron stars and heavy-ion-collision
experiments. In neutron stars, $\mu_I\neq0$ because of electric
charge neutrality. In heavy-ion-collision experiments, the initial
state corresponds to $\mu_B$, $\mu_I\neq0$ and the interaction time
is so short that both baryon number and isospin are conserved: the
electroweak interactions are irrelevant, and the strong interactions
determine the fate of the system. It is therefore phenomenologically
important to study the more general problem where both baryon and
isospin chemical potentials differ from zero.  Such conditions are
also obviously interesting on the theoretical perspective and a few
models have been used to study the strong interaction at nonzero
temperature, baryon and isospin chemical potentials \cite{muBIold,
muBInew, muBINc}.

The results obtained in some of these models are quite striking: the
phase diagram in the $(\mu_B,T)$ plane is qualitatively altered by
the introduction of a small $\mu_I$ \cite{muBInew, muBINc}. In
particular at small $\mu_B$ and high $T$, there are two phase
transitions or crossovers between the hadronic phase and the
quark-gluon-plasma phase when $\mu_I$ is small.  For fixed $\mu_B$
and $\mu_I$, the low temperature phase is the usual hadronic phase
where both the up and down quark condensates,
$\langle\bar{u}u\rangle$ and $\langle\bar{d}d\rangle$, are large but
different. When the temperature is increased, one of the condensates
becomes small, whereas the other one remains large. Finally, if the
temperature is further increased, both condensates become small. The
existence of these different phases is possible since $\mu_B$,
$\mu_I\neq0$ corresponds to $\mu_u\neq\mu_d$, where $\mu_{u,d}$ are
the quark chemical potentials for the up and down quark flavors, and
there is no symmetry that enforces $\langle\bar{u}u\rangle =
\langle\bar{d}d\rangle$ even when the light quark flavors have the
same mass.

However, the axial anomaly, which results in quark flavor mixing,
might invalidate these results \cite{muBInew, muBINc}.  Indeed, if
the axial anomaly is important enough, it is not possible for one of
the light quark condensates to be large while the other one is
small. As a consequence, for a large enough axial anomaly, a small
isospin chemical potential has little consequences for the phase
diagram. In this article, we shall use a simple model based on
chiral perturbation theory and the virial expansion to determine
whether our current phenomenological understanding of the strong
interaction in the hadronic phase simultaneously allows for a
significant difference between the two light quark condensates at
given temperature, baryon and isospin chemical potentials. Our
phenomenology-based analysis provides a test whether the axial
anomaly is strong enough to enforce
$\langle\bar{u}u\rangle\simeq\langle\bar{d}d\rangle$.  We limit
ourselves to $|\mu_I|<m_\pi$, where $m_\pi$ is the pion mass, in
order to avoid pion superfluid phases \cite{superfluid,
lattMuReview}. We first describe our method and then evaluate
quantitatively the difference between $\langle\bar{u}u\rangle$ and
$\langle\bar{d}d\rangle$ at nonzero temperature, baryon and isospin
chemical potentials.

\section{Light Quark Condensates}

In the hadronic phase, at low enough temperatures and chemical
potentials, the physics of the strong interaction is dominated by
the pseudo-Goldstone bosons due to the spontaneous breaking of
chiral symmetry since they are the lightest excitations in the
spectrum.  This fact is at the basis of chiral perturbation theory
\cite{chpt}. At finite temperature, it has been shown that the
pseudo-Goldstone bosons are the most important modes below
$\sim150$~MeV \cite{GerberL}.  Above this temperature, massive modes
have to be taken into account:  although exponentially suppressed,
their role starts to be significant.

The temperature dependence of the light quark condensates at zero
chemical potentials has been thoroughly studied in \cite{GerberL} up
to three loops in chiral perturbation theory and included the
contribution due to the massive modes.  In this case, and for equal
light quark masses, $m_u=m_d$, the two corresponding condensates,
$\langle\bar{u}u\rangle$ and $\langle\bar{d}d\rangle$, are equal.
However at nonzero baryon and isospin chemical potentials, the
symmetry that enforces
$\langle\bar{u}u\rangle=\langle\bar{d}d\rangle$ is explicitly broken
since $\mu_u\neq\mu_d$.  In this article, we use a method similar to
\cite{GerberL} to show that the difference between the light quark
condensates can become significant at large enough temperature and
chemical potentials.

The QCD partition function at nonzero quark chemical potentials is
given by
\begin{eqnarray}
Z_{\rm QCD}=\int dA  \; e^{-S_{\rm YM}} \; \prod_f {\rm
det}(i\Dirac+m_f+\mu_f \gamma_0),
\end{eqnarray}
where $m_f$ and $\mu_f$ are the mass and the chemical potential
related to the quark flavor $f$.  In this article we will consider
the case $m_u=m_d\neq m_s$, $\mu_u\neq\mu_d$, and $\mu_s=0$.  The
quark condensate for the flavor $f$ is given by
\begin{eqnarray}
\langle\bar{q}_fq_f\rangle=\lim_{V\rightarrow\infty} \frac TV
\frac{\partial}{\partial m_f} \ln Z_{\rm QCD}.
\end{eqnarray}

At low temperatures, the dominant contribution to the light quark
condensates comes from the pseudo-Goldstone bosons.  The physics of
these mesons is described by the chiral perturbation theory
partition function, $Z_{\rm ChPT}$.  However, since chiral
perturbation theory contains only mesons, $Z_{\rm ChPT}$ at
$\mu_s=0$ depends only on the isospin chemical potential,
$\mu_I=\mu_d-\mu_u$, not on the baryon chemical potential
$\mu_B=\frac32(\mu_u+\mu_d)$. Notice that for $\mu_s=0$, the
strangeness chemical potential $\mu_S=-\frac13\mu_B$.  Therefore,
for equal light quark masses, $m_u=m_d$, the pseudo-Goldstone bosons
contribute equally to $\langle\bar{u}u\rangle$ and
$\langle\bar{d}d\rangle$, since $Z_{\rm ChPT}(\mu_I)=Z_{\rm
ChPT}(-\mu_I)$. We thus conclude that in the hadronic phase, the
difference between the two light quark condensates comes solely from
the massive modes. In the hadronic phase, the contribution of a
hadron of mass $m_H$ is exponentially suppressed by the Boltzmann
factor $\sim\exp(-m_H/T)$, and its interaction with another hadron
of mass $m_H'$ is damped by a factor $\sim \exp(-(m_H+m_H')/T)$.
Interactions between hadrons and pions are further suppressed
because pions are pseudo-Goldstone modes and therefore interact
weakly at rest. In the hadronic phase, where temperatures never
exceed $\sim200$~MeV, it should therefore be sufficient to treat the
massive modes in the free gas approximation.

In the free gas approximation, the difference between the light
quark condensates is given by
\begin{eqnarray}
\label{Delta}
\delta=\frac{\langle\bar{u}u-\bar{d}d\rangle}{\langle0|\bar{q}q|0\rangle}
=-\sum_H \frac{g_H}{\langle0|\bar{q}q|0\rangle} \;
\sqrt{\frac{m_HT^3}{8\pi^3}} \left( \frac{\partial m_H^2}{\partial
m_u} - \frac{\partial m_H^2}{\partial m_d} \right) e^{-m_H/T} \;
\cosh\left( \frac{(B_H+\frac13S_H)\mu_B-I_{H3}\mu_I}T \right),
\end{eqnarray}
where
$\langle0|\bar{q}q|0\rangle=\langle0|\bar{u}u|0\rangle=\langle0|\bar{d}d|0\rangle$
is the light quark condensate at zero temperature and chemical
potentials. The sum in the equation above is over massive hadrons,
$H$, with spin degeneracy $g_H$, mass $m_H$, baryon number $B_H$,
third component of isospin $I_{H3}$, and strangeness $S_H$.  We have
used that $\mu_S=-\frac13\mu_B$ when $\mu_s=0$.

We now have to determine $\partial m_H^2/\partial m_f$.  Any hadron
mass can be written as
\begin{eqnarray}
\label{mH} m^2_H=\bar{m}^2_H+C_H I_{H3} (m_u-m_d),
\end{eqnarray}
where $\bar{m}_H$ is the hadron mass at $m_u=m_d$ and includes the
contributions of the strange and heavier quarks. Since the up and
down quarks are light, it is sufficient to use a linear expansion in
(\ref{mH}): the linear term dominates over other terms that depend
on $m_u-m_d$. The quantity $C_H$ comes from isovector interactions
and is related to the energy cost to change the third component of
isospin by one unit, i.e. to replace a $u$ quark by a $d$ quark,
within an isospin multiplet. Therefore we find that
\begin{eqnarray}
\label{derivH} \frac{\partial m_H^2}{\partial m_u} - \frac{\partial
m_H^2}{\partial m_d}=2 C_H I_{H3},
\end{eqnarray}
and the difference between the light quark condensates reads
\begin{eqnarray}
\label{Delta2} \delta= -\sum_H \frac{C_H I_{H3}
g_H}{\langle0|\bar{q}q|0\rangle} \; \sqrt{\frac{m_HT^3}{2\pi^3}}
e^{-m_H/T} \; \cosh\left( \frac{(B_H+\frac13S_H)\mu_B-I_{H3}\mu_I}T
\right).
\end{eqnarray}
Equivalently, if we sum over each isospin multiplet we get that
\begin{eqnarray}
\label{DeltaMultiplet} \delta=\sum_{{\rm isospin \; multiplets}}
\frac{C_H g_H}{\langle0|\bar{q}q|0\rangle} \; \sqrt{\frac{2
m_HT^3}{\pi^3}} e^{-m_H/T} \; \sinh((B_H+\frac13S_H)\mu_B/T)
\sum_{I_{H3}\geq0} I_{H3} \sinh(I_{H3}\mu_I/T),
\end{eqnarray}
where we have used that isospin symmetry at $m_u=m_d$ implies that
$m_H$ and $C_H$ do not change inside an isospin multiplet.
Therefore, since on general grounds $C_H>0$, we have shown that the
light quark condensates do differ in the hadronic phase at nonzero
temperature, baryon and isospin chemical potentials. Notice that
(\ref{DeltaMultiplet}) implies that only baryons with nonzero third
component of isospin contribute to the difference between the light
quark condensates.

\section{Quantitative Results}
In this section, we evaluate the importance of the difference
between the light quark condensates at nonzero temperature, baryon
and isospin chemical potentials.  It is clear from the above
analysis that the quantitative difference between the condensates
depends on the size of $C_H$, and that this difference will grow
exponentially when either the temperature, the baryon or the isospin
chemical potentials are increased.  The three lightest isospin
multiplets that contribute to $\delta$ are the $N(939)$, the
$\Sigma(1193)$, and the $\Delta(1232)$.

For the nucleon, the Feynman-Hellman theorem implies that $C_N=2 m_N
\langle p|\bar{u}u-\bar{d}d|p\rangle$. This can be estimated by
using $SU(3)$ flavor symmetry \cite{uMdsu3}:
\begin{eqnarray}
\label{CN}
\frac{C_N}{\langle0|\bar{q}q|0\rangle}\simeq\frac{m_\Xi^2-m_\Sigma^2}
{(m_s-\hat{m})\langle0|\bar{q}q|0\rangle}
\simeq\frac{2(m_\Xi^2-m_\Sigma^2)}{(\frac{m_s}{\hat{m}}-1) m_\pi^2
F_\pi^2},
\end{eqnarray}
where $\hat{m}=(m_u+m_d)/2$ and we have used the
Gell-Mann--Oakes--Renner relation: $m_\pi^2 F_\pi^2=2
\hat{m}\langle0|\bar{q}q|0\rangle$.  Numerically, if we use that
$m_s\simeq25 \hat{m}$ \cite{LeutwylerMass}, we get that $C_N/
\langle0|\bar{q}q|0\rangle \simeq (1.3 \; 10^{-2}$~MeV$^{-1})^2$.
Much less is known about $C_\Sigma$ and $C_{\Delta}$.  We shall
assume that $C_{N}\simeq C_{\Sigma} \simeq C_{\Delta}$, which is
true in the large $N_c$ limit. The results depicted in Fig.~1 were
produced using these values for $C_{H}$ in (\ref{DeltaMultiplet})
with the $N$, $\Sigma$ and $\Delta$ isospin multiplets.

\vspace{.75cm}
\begin{figure}[h]
\hspace{-.8cm}
\includegraphics[scale=0.33, clip=true, angle=0,draft=false]{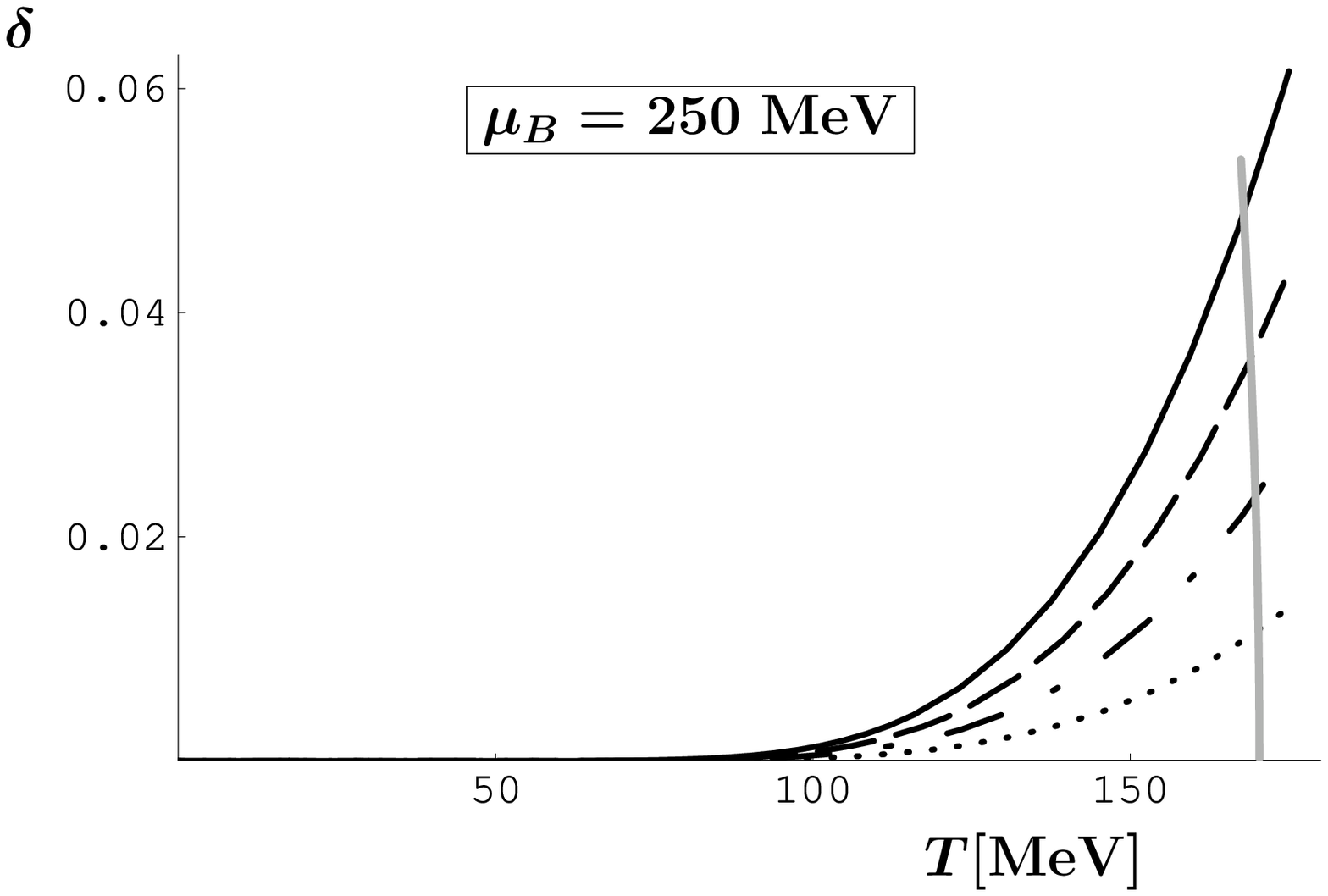}
\hspace{.5cm}
\includegraphics[scale=0.33, clip=true,
angle=0,draft=false]{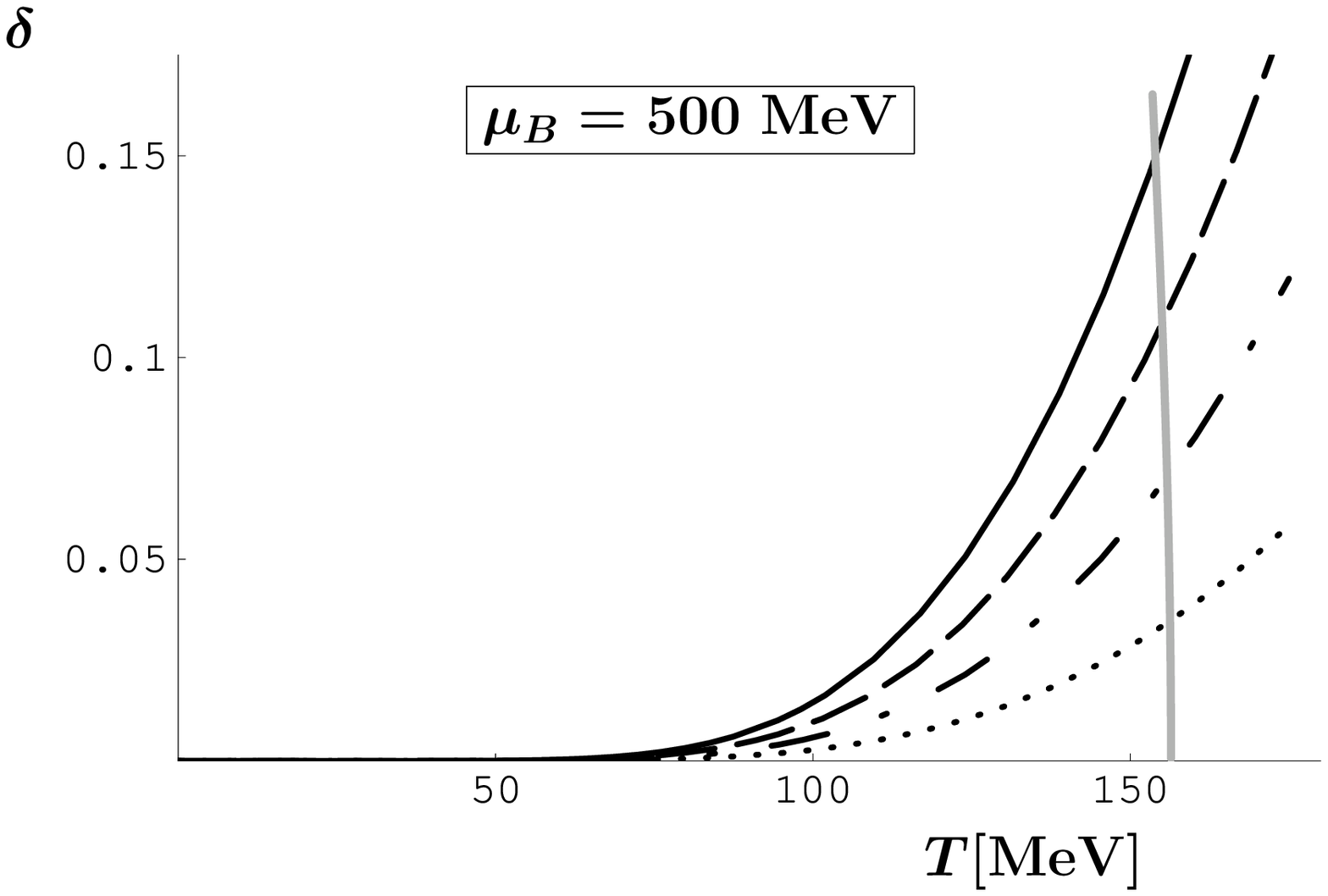} \hspace{.5cm}
\includegraphics[scale=0.33, clip=true,
angle=0,draft=false]{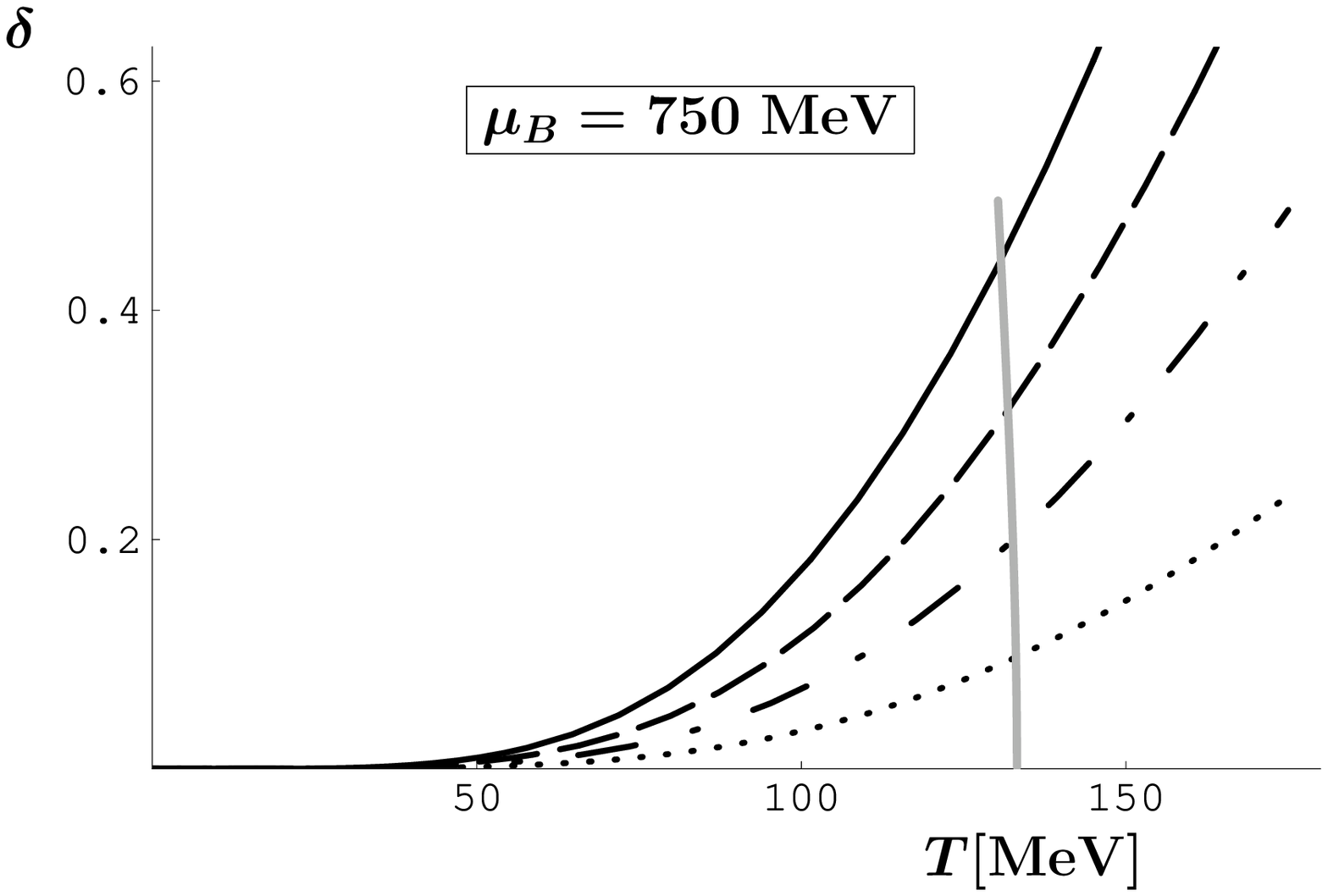} \caption{The difference between
the up and down quark condensates,
$\delta=\langle\bar{u}u-\bar{d}d\rangle/\langle0|\bar{q}q|0\rangle$
as a function of temperature at fixed baryon and isospin chemical
potentials. The baryon chemical potential is set to $250$, $500$,
and $750$~MeV in the first, second, and third graphs, respectively.
In each graph, the dotted line corresponds to $\mu_I=30$~MeV, the
dash-dotted line to $\mu_I=60$~MeV, the dashed line to
$\mu_I=90$~MeV, and the solid line to $\mu_I=120$~MeV. The light
gray line corresponds to the critical temperature determined by the
peak of a flavor insensitive susceptibility \cite{lattMuReview}.}
\end{figure}
\vspace{.3cm}

As shown in Fig.~1, we find that the difference between the up and
down quark condensates can be relatively large compared to the value
of these condensates in the vacuum: up to $\sim50$\% for
$\mu_B=750$~MeV and $\mu_I=120$~MeV.  Of course, these are quite
large chemical potentials, but our study demonstrates that, as a
matter of principle, there can be a sizeable difference between
$\langle\bar{u}u\rangle$ and $\langle\bar{d}d\rangle$ at nonzero
baryon and isospin chemical potentials.  The size of the difference
depends mainly on $C_{H}$.  The uncertainty on this quantity is
rather large, and our quantitative results cannot be very precise.
At large enough $\mu_I$, the largest contribution to $\delta$ comes
from $\Delta(1232)$, which dominates over the nucleon because of its
larger third component of isospin and its larger spin.  In
comparison, the contribution of the $\Delta(1600)$ resonance is
negligible compared to that of the nucleon.  To check the validity
of the free gas approximation we have calculated the next-to-leading
term in the virial expansion taking into account pion-nucleon
interactions.  We have found that this term is negligible compared
to the leading term in the virial expansion, i.e. the free gas
approximation. This stems from the fact that pions are
pseudo-Goldstone bosons and thus are weakly interacting particles.
In \cite{hgrLatt}, it was shown that heavier resonances can have a
significant impact when taken collectively into account.  In our
case, we only used the lightest contributing modes in the spectrum
to evaluate $\delta$. Heavier baryons with nonzero third component
of isospin will also contribute to $\delta$, with the same sign.
Therefore, the contributions from heavier resonances will add up and
increase $\delta$ for any temperature, baryon and isospin chemical
potentials.  However, because of the lack of precision on the value
of $C_{H}$ for heavier baryons, we decided not to evaluate their
contributions and restrict ourselves to the three lightest isospin
multiplets which should represent the dominant contribution.

\section{Conclusions and Outlook}
We have shown that the up and down quark condensates can
significantly differ at nonzero temperature, baryon and isospin
chemical potentials.  We have evaluated their difference in a simple
model. We have found that the difference between the light quark
condensates increases exponentially with the temperature, baryon or
isospin chemical potentials.  This difference becomes significant
for large enough chemical potentials.

These results imply that there are two phase transitions or
crossovers between the hadronic phase and the quark-gluon-plasma
phase at nonzero baryon and isospin chemical potentials, as
different models have predicted \cite{muBInew, muBINc}. In the
hadronic phase, both $\langle\bar{u}u\rangle$ and
$\langle\bar{d}d\rangle$ are large but different.  In the
quark-gluon-plasma phase, both light quark condensates are very
small.  Our results show that, at nonzero baryon and isospin
chemical potentials, there is an intermediate phase where only one
of the light quark condensates is large while the other one is very
small.  As has been discussed in \cite{muBInew,muBINc}, the
existence of this intermediate phase leads to important qualitative
changes for the QCD phase diagram and for the nature of the critical
endpoint, which might have interesting signatures in
heavy-ion-collision experiments.

\begin{acknowledgments}
It is a pleasure to thank T.~Cohen and J.~Gasser for useful
discussions. The Particle Physics Group at the Rensselaer
Polytechnic Institute, where part of this work was completed, is
also thanked for its warm hospitality. Work supported by the NSF
under grant No.~NSF-PHY0304252.
\end{acknowledgments}

\end{document}